\documentstyle[12pt,fullpage]{article}
\begin{document}
\begin{center}
{\Large\bf Charged-Surface Instability Development in Liquid Helium;
Exact Solutions}

\bigskip
{\large\bf N. M. Zubarev}

\bigskip
{\small\sl Institute of Electrophysics, Ural Division, Russian Academy of
Sciences,\\
106 Amundsena Street, 620016 Ekaterinburg, Russia\\
e-mail: nick@ami.uran.ru}
\end{center}

\begin{quotation}
{\small
The nonlinear dynamics of charged-surface instability development was
investigated for liquid helium far above the critical point. It is found
that, if the surface charge completely screens the field above the
surface, the equations of three-dimensional (3D) potential motion of a
fluid are reduced to the well-known equations describing the 3D Laplacian
growth process. The integrability of these equations in 2D geometry allows
the analytic description of the free-surface evolution up to the formation
of cuspidal singularities at the surface.
}
\end{quotation}

\medskip
It is known [1] that the flat electron-charged surface of liquid helium is
unstable if the electric field strength above $(E_{+})$ and inside
$(E_{-})$ the fluid satisfy inequality
$$
{E_+}^2+{E_-}^2>{E_c}^2=8\pi\sqrt{g\alpha\rho},
$$
where $g$ is the free fall acceleration, $\alpha$ is the surface tension
coefficient, and $\rho$ is the fluid density. An analysis of the critical
behavior of the system suggests that, depending on the dimensionless
parameter $S=({E_-}^2-{E_+}^2)/{E_c}^2$, the nonlinearity leads either to
the saturation of linear instability or, conversely, to the explosive
increase in amplitude. The first situation may result in the formation of
a stationary perturbed surface relief (hexagons [2] and many-electron
dimples [3]) in liquid helium. The use of a perturbation theory, with the
surface slope as a small parameter, allowed the detailed analytic study of
such structures in the critical region [4,5]. In the second case, the
small-angle approximation fails. The cinematographic study by V.P.
Volodin, M.S. Khaikin, and V.S. Edelman [6] has demonstrated that the
development of surface instability leads to the formation of dimples and
their sharpening in a finite time. A substantial
nonlinearity of this processes calls for a theoretical model that is free
from the requirement for smallness of surface perturbations and adequately
describes the formation dynamics of a singular surface profile in liquid
helium. This work demonstrates that such a model can be developed if the
condition $E_-\gg E_+$ is fulfilled, i.e., the field above liquid helium
is fully screened by the surface electron charge, and if the electric
field far exceeds its critical value, i.e., $E_-\gg E_c$.

Let us consider the potential motion of an ideal fluid (liquid helium), in
a region bounded by the free surface $z=\eta(x,y,t)$. We assume that the
characteristic scale $\lambda$ of surface perturbations is much smaller
than the fluid depth. We also assume that
$$
\alpha E_-^{-2}\ll\lambda\ll{E_-}^2/(g\rho),
$$
so that the capillary and gravity effects can be ignored. The
electric-field potential $\varphi(x,y,z,t)$ in the medium and the fluid
velocity potential $\Phi(x,y,z,t)$ satisfy Laplace equations
\begin{equation}
\nabla^2\varphi=0, \qquad
\nabla^2\Phi=0,
\end{equation}
which should be solved jointly with the conditions
at the surface
\begin{equation}
8\pi\rho\Phi_t+4\pi\rho(\nabla\Phi)^2+(\nabla\varphi)^2={E_-}^2,
\qquad z=\eta(x,y,t),
\end{equation}
\begin{equation}
\eta_t=\Phi_z-\nabla_{\!\!\bot}\eta\cdot\nabla_{\!\!\bot}\Phi,
\qquad z=\eta(x,y,t),
\end{equation}
\begin{equation}
\varphi=0, \qquad z=\eta(x,y,t),
\end{equation}
and conditions at infinity
\begin{equation}
\varphi\to -zE_-, \qquad  z\to-\infty,
\end{equation}
\begin{equation}
\Phi\to 0, \qquad z\to-\infty.
\end{equation}

Let us pass to the dimensionless variables, taking $\lambda$ as a length
unit, $E_-$ as a unit of electric field strength, and
$\lambda E_-^{-1}(4\pi\rho)^{1/2}$ as a time unit. It is convenient to
rewrite the equations of motion of free surface $z=\eta(x,y,t)$ in the
implicit form (not containing the $\eta$ function explicitly). Let us
introduce the perturbed potential $\tilde{\varphi}=\varphi+z$ decaying at
infinity. One has at the boundary: $\tilde{\varphi}|_{z=\eta}=\eta$. It is
then straightforward to obtain the following relationships:
$$
\eta_t=\left.\frac{\tilde{\varphi}_t}
{1-\tilde{\varphi}_z}\right|_{z=\eta},
\qquad
\nabla_{\!\!\bot}\eta=\left.\frac{\nabla_{\!\!\bot}\tilde{\varphi}}
{1-\tilde{\varphi}_z}\right|_{z=\eta},
$$
which allow one to eliminate the $\eta$ function from Eq. (3). The
dynamic and kinematic boundary conditions (2) and (3) then take the form
\begin{equation}
\Phi_t-\tilde{\varphi}_z=-(\nabla\Phi)^2/2-(\nabla\tilde{\varphi})^2/2,
\qquad z=\eta(x,y,t),
\end{equation}
\begin{equation}
\tilde{\varphi}_t-\Phi_z=-\nabla\tilde{\varphi}\cdot\nabla\Phi,
\qquad z=\eta(x,y,t).
\end{equation}
Let now introduce a pair of auxiliary potentials:
$$
\phi^{(\pm)}(x,y,z,t)=(\tilde{\varphi}\pm\Phi)/2.
$$
With these potentials, the fluid surface shape can be defined by
relationship
\begin{equation}
\eta=\left.(\phi^{(+)}+\phi^{(-)})\right|_{z=\eta},
\end{equation}
while equations of motion (1)--(6) are reduced to the following symmetric
form:
\begin{equation}
\nabla^2\phi^{(\pm)}=0,
\end{equation}
\begin{equation}
\phi^{(\pm)}_t=\pm\phi^{(\pm)}_z\mp(\nabla\phi^{(\pm)})^2,
\qquad z=\eta(x,y,t),
\end{equation}
\begin{equation}
\phi^{(\pm)}\to 0, \qquad z\to-\infty,
\end{equation}
where boundary conditions (11) are obtained by combining Eqs. (7) and (8) with
plus and minus sign, respectively.

It is seen that the equations of motion are split into two systems
of equations for the potentials $\phi^{(+)}$ and $\phi^{(-)}$, which are
implicitly related by equation for surface shape (9). An essential point
is that these equations are compatible with either $\phi^{(-)}=0$ or
$\phi^{(+)}=0$ condition. One can readily see that the first condition
corresponds to those solutions whose amplitude increases with time, while
the second condition corresponds to the decaying solutions that are of no
interest to us.

Thus, an analysis of the equations of motion of a charged surface of
liquid helium reveals a solution increasing with $t$ and corresponding to
the $\phi^{(-)}=0$ condition or, what is the same, to the $\varphi+z=\Phi$
condition (the stability of this branch of solutions is proved below).
The functional relation between the potentials can be used to eliminate
the velocity potential $\Phi$ from initial Eqs. (1)--(6). In the moving
system of coordinates $\{x',y',z'\}=\{x,y,z-t\}$, one has
\begin{equation}
\nabla^2\varphi=0,
\end{equation}
\begin{equation}
\eta'_t=\partial_n\varphi\,\sqrt{1+(\nabla_{\!\!\bot}\eta')^2},
\qquad z'=\eta'(x',y',t).
\end{equation}
\begin{equation}
\varphi=0, \qquad z'=\eta'(x',y',t)
\end{equation}
\begin{equation}
\varphi\to -z', \qquad z'\to-\infty,
\end{equation}
where $\eta'(x',y',t)=\eta-t$, and $\partial_n$ denotes the normal
derivative. These equations explicitly describe the motion of a free
charged surface $z'=\eta'(x',y',t)$. They coincide with the equations for
the so-called Laplacian growth process, i.e., the phase boundary movement
with velocity directly proportional to the normal derivative of a certain
scalar field ($\varphi$ in our case). Depending on the system, this may be
the temperature (Stefan problem in the quasi-stationary limit), the
electrostatic potential (electrolytic deposition), the pressure (flow
through a porous medium), etc.

Note that the boundary movement described by Eqs. (13)--(16) is invariably
directed inward from the surface. Let $\eta'$ be a single-valued function
of variables $x'$ and $y'$ at zero time $t=0$. Then the inequality
$\eta'(x',y',t)\leq\eta'(x',y',0)$ holds for $t>0$. In the initial
notations,
\begin{equation}
\eta(x,y,t)\leq\eta(x,y,0)+t
\end{equation}
for any $x$ and $y$. This condition can be used to prove the stability of
the ascending branch to the small perturbations of potential $\phi^{(-)}$.
Clearly, the boundary motion at small $\phi^{(-)}$ values is
entirely controlled by the potential $\phi^{(+)}$ [one should set
$\phi^{(-)}=0$ in Eq. (9)] and, hence, obeys Eqs. (13)--(16). The
evolution of the $\phi^{(-)}$ potential is described by Eqs. (10)--(12),
with the following simple boundary condition in the linear approximation:
$$
\phi^{(-)}_t=-\phi^{(-)}_z, \qquad z=\eta(x,y,t).
$$
Let the potential distribution at zero time $t=0$ be determined by the
expression
$$
\phi^{(-)}|_{t=0}=\phi_0(x,y,z),
$$
where $\phi_0$ is a harmonic function at $z\leq\eta(x,y,0)$ decaying at
$z\to-\infty$. It is then straightforward to show that the time dynamics
of the $\phi^{(-)}$ potential is given by
$$
\phi^{(-)}=\phi_0(x,y,z-t).
$$
This implies that the singularities of the $\phi^{(-)}$ function will
drift in the $z$ direction, so that they will occur only in the
$z>\eta(x,y,0)+t$ region. Taking into account inequality (17), one finds
that the singularities always move away from the boundary $z=\eta(x,y,t)$
of liquid helium. Consequently, the perturbation $\phi^{(-)}$ will relax
to zero, as we wished to prove.

Let us now turn to the analysis of the dynamics of surface instability
development in liquid helium. In the 2D case (all quantities are taken to
be independent of the $y$ variable), system of Eqs. (13)--(16) is reduced
to the well-known Laplacian growth equation (see, e.g., [7] and references
therein):
$$
\mbox{Im}(f^*_t f_w)=1, \qquad \varphi=0.
$$
In this expression, $f=x'+iz'$ is a complex function analytical in the
lower half-plane of the complex variable $w=\psi-i\varphi$ and satisfying
condition $f\to w$ at $w\to\psi-i\infty$. Note that the $\psi$ function is
a harmonic conjugate to $\varphi$, while the condition $\psi=\mbox{const}$
defines the electric field lines in a medium. The Laplacian growth
equation is integrable in the sense that it allows for the infinite number
of partial solutions:
$$
f(w)=w-it-i\sum_{n=1}^N a_n\ln\left(w-w_n(t)\right)
+i\left(\sum_{n=1}^N a_n\right)\ln\left(w-\tilde{w}(t)\right),
$$
where $a_n$ are complex constants and $\mbox{Im}(w_n)>0$. The last term is
added in order that the condition $\eta\to 0$ be fulfilled at
$|x|\to\infty$. One can put $\mbox{Im}(\tilde{w})\gg\mbox{Im}(w_n)$; in
this case, the influence of this term on the surface evolution can be
ignored. The functions $w_n(t)$ are defined by the following set of
transcendental equations [7]:
$$
w_n+it+i\sum_{m=1}^Na^*_m\ln\left(w_n-w^*_m\right)=C_n,
$$
where $C_n$ are arbitrary complex constants.

Let us consider the simplest ($N=1$) solution to the Laplacian growth
equation:
\begin{equation}
f(w)=w-it+i\ln(w-ir(t)), \qquad
r(t)-\ln r(t)=1+t_c-t,
\end{equation}
where $t_c$ is a real constant and the real function $r(t)\geq 1$. The
shape of a solitary perturbation corresponding to Eqs. (18) is given
parametrically by expressions
\begin{equation}
z(\psi,t)=\ln\sqrt{\psi^2+r^2(t)}, \qquad
x(\psi,t)=\psi-\arctan\left(\psi/r(t)\right).
\end{equation}
This solution exists only during a finite time period and culminates in the
formation, at time $t=t_c$, of a singularity in the form of a first-kind
cusp at the fluid surface. Indeed, setting $r=r(t_c)=1$ in Eq. (19), one
obtains
$$
2z=|3x|^{2/3}
$$
in the leading order near the singular point (see
also [8]). Note that the electric field turns to infinity at the cusp:
$$
\partial_n\varphi\sim\left.x_\psi^{-1}\right|_{\psi=0}\sim1/\sqrt{t_c-t}.
$$
The surface velocity also becomes infinite in a finite time:
$$
\eta_t=\left.z_t\right|_{\psi=0}\sim 1/\sqrt{t_c-t}.
$$
It is worth noting that the singular solution in the leading order is also
true when the field above the surface is screened incompletely. The point
is that the requirement that the field above the surface be small compared
to the field in fluid is naturally satisfied in the vicinity of the
singularity.

Let now discuss the influence of the capillary effects. One can readily
estimate the surface and electrostatic pressures near the surface:
$$
\alpha R^{-1}\sim\alpha\rho^{1/2}E_-^{-1}\,(t_c-t)^{-1},
\qquad
(\partial_n\varphi)^2\sim\lambda\rho^{1/2}E_-\,(t_c-t)^{-1}.
$$
Insofar as we assumed that $\lambda\gg\alpha E_-^{-2}$, the capillary
forces cannot compete with the electrostatic ones, so that there is no
need to take into account the surface forces at the stage of cusp
formation.

In summary, we succeeded in finding a broad class of exact solutions to
the equations of motion of a charged surface of liquid helium. It is
remarkable that the solutions obtained are not constrained by the
condition for the smallness of surface perturbations: the model suggested
describes the free-surface instability development up to the formation of
the singularities (cusps) similar to those observed in the experiment [6].

\medskip
I am grateful to E.A. Kuznetsov for stimulating discussions.
This work was supported by the Russian Foundation for Basic Research
(project no. 00-02-17428) and the INTAS (grant no. 99-1068).

\bigskip

{\small
\begin{enumerate}
\item L.P. Gorkov and D.M. Chernikova, Sov. Phys. Dokl. {\bf 21}, 328 (1976).
\item M. Wanner and P. Leiderer. Phys. Rev. Lett. {\bf 42}, 315 (1979).
\item A.A. Levchenko, E. Teske, G.V. Kolmakov, {\sl et al.}, JETP Lett.
{\bf 65}, 572 (1997).
\item V.B. Shikin and P. Leiderer, JETP Lett. {\bf 32}, 572 (1980).
\item V.I. Melnikov and S.V. Meshkov, Sov. Phys. JETP {\bf 54}, 505 (1981);
{\bf 55}, 1099 (1982).
\item V.P. Volodin, M.S. Khaikin, and V.S. Edelman, JETP Lett. {\bf 26}, 543
(1977).
\item M.B. Mineev-Weinstein and S.P. Dawson, Phys. Rev. E {\bf 50}, R24
(1994).
\item S.D. Howison, SIAM J. Appl. Math. {\bf 46}, 20 (1986).
\end{enumerate}
}

\end {document}